\documentclass[usegraphicx,usenatbib,natbib,useAMS]{mn2e}

\usepackage{times}     
\usepackage{aas_macros}
\usepackage{amssymb}   
\usepackage{multirow}

\def\ebv{\mbox{$E(B-V)$}}
\def\halpha{\mbox{H$\alpha$}}
\def\hbeta{\mbox{H$\beta$}}
\def\hgamma{\mbox{H$\gamma$}}
\def\hdelta{\mbox{H$\delta$}}
\def\lya{\mbox{Ly$\alpha$}}

\def\ecs{\mbox{\,erg~cm$^{-2}$~s$^{-1}$}}

\def\ecsa{\mbox{\,erg~cm$^{-2}$~s$^{-1}$~{\AA}$^{-1}$}}
\def\lesssim{\mathrel{\hbox{\rlap{\hbox{\lower4pt\hbox{$\sim$}}}\hbox{$<$}}}}
\def\gtrsim{\mathrel{\hbox{\rlap{\hbox{\lower4pt\hbox{$\sim$}}}\hbox{$>$}}}}

\newcommand{\kms}{\ensuremath{{\rm km\,s^{-1}}}}
\newcommand{\A}{A31.1} 

\DeclareRobustCommand{\ion}[2]{%
\relax\ifmmode
 \ifx\testbx\f@series
  {\mathbf{#1\,\mathsc{#2}}}\else
  {\mathrm{#1\,\mathsc{#2}}}\fi
 \else\textup{#1\,{\mdseries\textsc{#2}}}%
\fi}

\title[Abundances of Lensed Galaxies]{Gravitationally Lensed Galaxies
  at $2<z<3.5$: Direct Abundance Measurements of \lya\ Emitters
  \thanks{Based on data from the X-shooter GTO observations collected
    at the European Southern Observatory VLT/Kuyuen telescope,
    Paranal, Chile, collected under programme IDs: 084.B-0351(D),
    086.A-0674(A), 086.A-0674(B), 087.A-0432(A) and 087.A-0432(B).}}

\author[L. Christensen et al.]
       {Lise Christensen\thanks{lise@dark-cosmology.dk}$^{1}$,
         Peter Laursen$^{2,1}$, 
         Johan Richard$^{3,1}$, 
         Jens Hjorth$^1$, 
         Bo Milvang-Jensen$^1$, 
         \newauthor         
         Miroslava Dessauges-Zavadsky$^4$, 
         Marceau Limousin$^5$,
         Claudio Grillo$^{1}$, 
         Harald Ebeling$^6$\\
        $^1$ Dark Cosmology Centre, Niels Bohr Institute, University of Copenhagen, Juliane Maries Vej 30, 2100 Copenhagen, Denmark\\
        $^2$ Oskar Klein Centre, Dept. of Astronomy, Stockholm University, SE-10691 AlbaNova, Stockholm, Sweden\\
        $^3$ Centre de Recherche Astrophysique de Lyon, Universit´e Lyon 1, Observatoire de Lyon, 9 Avenue Charles Andr´e, 69561 Saint Genis
Laval cedex, France\\
        $^4$ Observatoire de Gen\`eve, Universit\'e de Gen\`eve, 51 Ch. des Maillettes, 1290 Sauverny, Switzerland\\
        $^5$  Aix Marseille Universit\'e, CNRS, LAM (Laboratoire d'Astrophysique de Marseille) UMR 7326, 13388, Marseille, France\\
        $^6$ Institute for Astronomy, University of Hawaii, 2680 Woodlawn Drive, Honolulu, HI 96822, USA\\
}
\date{Accepted 2012 August 29. Received 2012 August 28}            
\pubyear{2012}

\begin{document}

\maketitle

\label{firstpage}

\begin{abstract}
Strong gravitational lensing magnifies the flux from distant galaxies,
allowing us to detect emission lines that would otherwise fall below
the detection threshold for medium-resolution spectroscopy. Here we
present the detection of temperature-sensitive oxygen emission lines
from three galaxies at $2\lesssim z\lesssim3.5$, which enables us to
directly determine the oxygen abundances and thereby double the number
of galaxies at $z>2$ for which this has been possible. The three
galaxies have $\sim$10\% solar oxygen abundances in agreement with
strong emission line diagnostics. Carbon and nitrogen ratios relative
to oxygen are sub-solar as expected for young metal-poor galaxies. Two
of the galaxies are \lya\ emitters with rest-frame equivalent widths
of 20 {\AA} and 40 {\AA}, respectively, and their high magnification factors
allow us for the first time to gain insight into the physical
characteristics of high-redshift \lya\ emitters. Using constraints
from the physical properties of the galaxies, we accurately reproduce
their line profiles with radiative transfer models. The models show a
relatively small outflow in agreement with the observed small velocity
offsets between nebular emission and interstellar absorption lines.
\end{abstract}


\section{Introduction}

Strong gravitational lensing of distant galaxies behind massive galaxy
clusters allows us to study intrinsically fainter galaxies than
possible for regular field galaxies. With magnification factors
reaching up to 50, it becomes possible to study the detailed physical
properties of galaxies that are fainter than the characteristic
luminosity, $L^*$, at a given redshift. Since the first detailed
analyses of a lensed Lyman break galaxy, MS~1512-cB58
\citep{pettini00,teplitz00,pettini02}, a steady increase in the number
of spectroscopic observations of bright galaxies at $z\sim2$ either
detected serendipitously or in large dedicated surveys have appeared
\citep[e.g.][]{fosbury03,cabanac05,swinbank07,allam07,stark08,lin09,diehl09,hainline09,bian10,rigby11}.
To date, most lensed galaxies that have received more attention still
belong to the bright end of the luminosity function, although recent
investigations have started to probe the lower-mass end of the galaxy
distribution \citep{richard11,wuyts12}. Some galaxies show
particularly strong UV emission lines, suggesting the presence of a
large population of very massive, hot stars
\citep{fosbury03,villar-martin04,raiter10}, giving rise to nebular
emission lines that are very rarely detected in nearby galaxies. Other
strongly star-forming lensed galaxies have UV spectra that do not
differ substantially from the cB58 galaxy spectrum
\citep{quider09,quider10,dessauges10,dessauges11}.

With the boost of the fluxes from gravitational lensing, it also
becomes feasible to detect emission lines such as [\ion{O}{iii}]
$\lambda$4363 or [\ion{N}{iii}] $\lambda$5755 which, when related to
other transitions from the same ionisation stage, are sensitive to the
gas temperature. Combining this with the knowledge of the density of
the gas, it is possible to derive the abundances of various chemical
elements directly \citep{osterbrock89}. This approach is widely used
to determine abundances in low-redshift galaxies. The flux from the
[\ion{O}{iii}] $\lambda$4363 line increases with decreasing
metallicity; however, low-metallicity galaxies at high redshift are
also very faint, so the line is difficult to detect, in particular
since the line is redshifted to near-IR part of the spectrum.  The
high-redshift Lyman break galaxies (LBGs) with $<$25.5 mag
\citep{steidel96}) are known to be massive galaxies, and because of
the relation between mass and metallicity, LBGs are typically metal
rich \citep{erb06c}. This implies that their temperature-sensitive
lines have fluxes below the detection threshold even on the largest
telescopes. To date, only one detection of [\ion{O}{iii}]
$\lambda$4363 from a galaxy at $z>1$ has been reported \citep{yuan09},
while other attempts to measure the line have led to upper limits only
\citep{rigby11}.  Analysing other rest-frame UV lines (\ion{O}{iii}]
  $\lambda\lambda$1661,1666), which are also sensitive to the gas
  temperature may alleviate this difficulty of detecting
  [\ion{O}{iii}] $\lambda$4363 and determining abundances directly
  \citep{villar-martin04,erb10}. However, this doublet is also rarely
  observed in high-redshift galaxies, and it is more sensitive to
  uncertainties in the correction of extinction in the UV range of the
  spectrum.

To avoid the need for very long observation times to detect the
temperature sensitive lines, alternative methods have been
developed. Using only strong rest-frame optical emission lines,
various diagnostics have been calibrated with respect to the direct
abundance determinations
\citep[e.g.,][]{pagel79,pettini04,kewley08}. Since the
temperature-sensitive lines are generally too weak to be measured even
in bright emission-line galaxies at high redshifts, we are forced to
use the relations established at low redshifts to derive
abundances. The validity of using the exact same strong-line
diagnostics at high redshifts has to be verified.

To find potential UV bright high-redshift galaxies, we have studied
a sample of gravitationally lensed galaxies at $1<z<6$, based on
selecting high surface brightness regions in extended arcs. In a
companion paper \citep[][hereafter Paper I]{christensen12a}, we
present the sample and the continuum emission properties of the
galaxies. We determine stellar masses from spectral fitting, star
formation rates from emission lines, and oxygen abundances.  We also
demonstrate that the low-mass galaxies exhibit a large scatter
compared to the fundamental relation for star-forming galaxies
\citep{lara-lopez10,mannucci10} and provide an alternative calibration
valid for low-mass galaxies.

Out of the 12 galaxies in our original sample, three galaxies at
$1.8<z<3.5$ show temperature-sensitive lines. Two of these are also
strong \lya\ emitters, while the third galaxy shows \lya\ in
absorption. In this paper we present a more detailed analysis of the
nebular emission lines from these three galaxies. We present the data
in Section~\ref{sect:data}, and derive the direct abundances and
determine escape fractions in Section~\ref{sect:emphys}. In Section
\ref{sect:model_lya} we model the two \lya\ emission lines using
constraints from the physical properties of the galaxies.

\section{Observational Data}
\label{sect:data}
The data for the lensed galaxies were obtained with X-shooter on the
VLT during separate observing runs in 2010--2011. One of the science
drivers for X-shooter \citep{dodorico06,vernet11} is to determine
redshifts of faint single targets and analyse emission line galaxies
at $z>1$.  In previous papers, we demonstrated the ability of
X-shooter for the observations of lensed galaxies
\citep{dessauges10,pettini10,christensen10}. In Paper I, we describe
the target selection, observations, data reduction, and analysis of
the spectra.  In this paper, we focus on the analysis of the final
emission line fluxes from three galaxies after correcting for many
effects: Galactic extinction, intrinsic extinction, slit losses, and
lens magnification. One of the lensed galaxies is found in the
well-studied Abell 1689 cluster, while the remaining two are found in
two southern equivalents to the MACS clusters \citep{ebeling01} with
declinations $<$~--~40 deg (H. Ebeling, unpublished). This section
presents the three galaxies which we can investigate in further
detail.

\subsection{Abell 1689 arc ID 31.1}
The arc ID 31.1 is adopted from \citet{limousin07}. For this target, a
wealth of emission lines are observed as listed in
Table~\ref{tab:a9.1}.  From the strongest emission lines, which are
not affected by telluric absorption, and also excluding \lya\, we find
the redshift $z=1.8339\pm 0.0003$. The last column in
Table~\ref{tab:a9.1} lists the line fluxes after correcting for
intrinsic extinction as described in Section~\ref{sect:ebv_gas}.

The \lya\ line profile is double-peaked with a much brighter red than
blue component, which is the characteristic signature of resonance
scattering of photons in an outflowing medium (see
Fig. ~\ref{fig:Lyafit}). Its rest-frame equivalent width is 40{\AA}. In Sect.~\ref{sect:model_lya} we model the line profile taking
into account the constraints from physical parameters determined from
other emission lines.

We detect the temperature-sensitive [\ion{O}{iii]} $\lambda$4363 line.
Its strength increases relative to [\ion{O}{iii]} $\lambda5007$ with
decreasing metallicity, but at high redshifts, the galaxies that have
the best signal-to-noise ratio spectra are also the brightest and most
massive, and hence most metal-rich ones at those redshifts.  To date,
[\ion{O}{iii]} $\lambda$4363 has only been observed in another lensed
galaxy at $z=1.7$, also behind the Abell 1689 cluster \citep{yuan09}.

\begin{figure}
\begin{center}
\includegraphics[bb=25 275 643 677, clip, width=8.5cm]{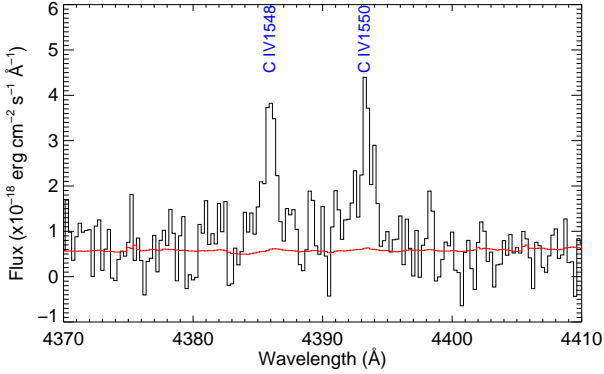}
\end{center}
\caption{Section of the observed UVB spectrum of the ID 31.1 arc
  showing its highly uncommon narrow \ion{C}{iv}
  $\lambda\lambda$1548,1550 emission lines. This suggest an ISM with a
  low covering factor along the line of sight.}
\label{fig:A31.1_CIV}
\end{figure}

Many other rest-frame UV emission lines are identified in the Abell
1689 ID 31.1 spectrum. In particular we draw attention to the narrow
emission lines from the \ion{C}{iv} $\lambda\lambda$1548,1550 doublet
shown in Fig.~\ref{fig:A31.1_CIV}. \ion{C}{iv}
$\lambda\lambda$1548,1550 are almost always seen in absorption in
high-redshift galaxies \citep[another exception is the Lynx arc
  in][]{holden01,fosbury03}, or in P-Cygni profiles with a minor
contribution from the emission component compared to the absorption
\citep{pettini00}.  This indicates a highly unobscured line of sight
to the galaxy. Searching for the UV absorption lines in the low
signal-to-noise ratio spectrum, we only detect weakly the \ion{Si}{ii}
$\lambda$1260,1264, and 1526 lines at $z=1.8348\pm0.0012$, while we do
not detect the typically strong high-ionisation absorption doublet
\ion{Si}{iv} $\lambda$1393,1402. Due to the weak absorption lines, we
argue that the interstellar medium (ISM) only has a partial coverage,
like observed for a couple of other lensed galaxies
\citep{quider09,dessauges10}. Among the uncommon emission lines, we
also detect the temperature-sensitive \ion{O}{iii]}
  $\lambda\lambda$1661,1666 lines, which among other high-$z$ galaxies
  are also observed in the bright $z=2.3$ galaxy Q2343--BX418
  \citep{erb10} and in the Lynx arc \citep{fosbury03,villar-martin04}.

\begin{table}
    \caption{Emission lines from the A1689 31.1 arc}
  \begin{tabular}{lccrr}
\hline
\hline
   line       &  $\lambda_r$ $^a$   & $\lambda_{\mathrm{obs}}$ &
   $f_{\mathrm{obs}}^b$ & $f_{\mathrm{cor}}^b$ \\
\hline
\lya          &   1215.67  & 3445.77        & 72.7$\pm$0.7& 4309$\pm$42\\
\ion{C}{iv}   &   1548.20  & 4386.81        & 4.3$\pm$0.4 & 137$\pm$13\\
\ion{C}{iv}   &   1550.77  & 4394.22        & 4.7$\pm$0.3 & 149$\pm$10\\
\ion{O}{iii]} &   1660.81  & 4705.82        & 3.2$\pm$0.4 & 90$\pm$11\\
\ion{O}{iii]} &   1666.15  & 4720.63        & 7.4$\pm$0.3 & 206$\pm$8\\
\ion{[C}{iii]}&   1906.68  & 5402.29        & 9.1$\pm$0.4 & 202$\pm$9\\
\ion{C}{iii]} &   1908.73  & 5408.19        & 4.7$\pm$0.4 & 104$\pm$9\\
\ion{[O}{ii]} &   3727.09  & 10563.06       & 9.2$\pm$0.8 & 68$\pm$6\\
\ion{[O}{ii]} &   3729.88  & 10569.40       &11.7$\pm$0.9 & 86$\pm$7\\
\ion{[Ne}{iii]}&  3869.84  & 10965.28       & 6.4$\pm$2.5 & 44$\pm$17\\  
\hgamma       &   4341.69  & 12301.90       &17.0$\pm$2.1 & 97$\pm$12\\
\ion{[O}{iii]}&   4364.44  & 12366.36       & 6.2$\pm$1.8 & 35$\pm$10\\
\hbeta        &   4862.70  & 13777.99$^d$   &42.7$\pm$4.9 & 205$\pm$24\\
\ion{[O}{iii]}&   4960.29  & 14057.55$^d$   &63.0$\pm$10.6& 293$\pm$49 \\
\ion{[O}{iii]}&   5008.24  & 14194.09$^{c}$ &211.5$\pm$5.7& 968$\pm$26\\
\hline        
\end{tabular}
\label{tab:a9.1}

\textit{Notes.~}\\ $^a$ Rest-frame vacuum wavelengths.\\ $^b$ Emission
line flux in units of $10^{-18}$ \ecs. $f_{\mathrm{obs}}$ and
$f_{\mathrm{cor}}$ correspond to the observed flux and that
corrected for intrinsic absorption, respectively.
\\ $^c$ Close to sky line.\\
$^d$ Affected by strong telluric absorption lines.
\end{table}

\subsection{SMACS\,J0304.3--4402 ID 1.1}

The X-shooter spectrum covers one of the multiple lensed images, and
within the slit three distinct regions are visible in the \emph{HST}
image (see Paper I). The emission lines listed in
Table~\ref{tab:m0304} present a sum of the total flux from the
components within the slit. The redshift for this galaxy is measured
to be $z=1.9634\pm0.0002$.  Again, we detect the \ion{O}{iii}]
  $\lambda\lambda$1661,1666 doublet. Apart from velocity offsets of
  $\sim$100 km s$^{-1}$ between emission lines from the two
  components, there are no significant differences in the relative
  line fluxes as verified by comparing extracted spectra from the two
  distinct component. The UV continuum has the same flux, but redwards
  of 2600 {\AA} in the rest-frame, the continuum emission differs by a
  factor of two, with the smallest flux coming from the most compact
  component. The emission line fluxes listed in Table~\ref{tab:m0304}
  are the total flux in the X-shooter slit summed for the two
  components. Even though the Balmer lines are very bright for this
  source, \lya\ emission is absent, and its (UV) spectrum appears
  similar to the cB58 spectrum \citep{pettini00}.

In addition to the strong UV absorption lines in the ISM, a strong
intervening \ion{Mg}{ii} at $z=1.5492$ is seen in the VIS spectrum
where its lines are blended with the strong \ion{Fe}{ii} lines of the
lensed galaxy's ISM.

\begin{table}
    \caption{Emission lines from the SMACS J0304 arc}
  \begin{tabular}{lccrr}
\hline
\hline
   line       &  $\lambda_r$ $^a$   & $\lambda_{\mathrm{obs}}$ &  $f_{\mathrm{obs}}^b$ &  $f_{\mathrm{cor}}^b$ \\
\hline

\ion{O}{iii]} &   1660.81  & 4921.93& 7.1$\pm$1.6 &   56$\pm$13\\
\ion{O}{iii]} &   1666.15  & 4937.35 & 11.4$\pm$1.9&   90$\pm$15\\
\ion{[O}{ii]}  &  2470.22  &7322.39& 8.4$\pm$2.1 &   45$\pm$11\\
\ion{[O}{ii]}  &  3727.09  &11044.78& 437.8$\pm$3.4&1514$\pm$12\\
\ion{[O}{ii]}  &  3729.88  &11053.09& 587.2$\pm$3.9&2029$\pm$14\\ 
H9             &  3836.49  &11370.55$^c$& 29.1$\pm$4.2 & 98$\pm$14\\
\ion{[Ne}{iii]}&  3869.84  &11468.93$^c$& 149.9$\pm$3.8& 499$\pm$13\\      
\ion{He}{i} + H8& 3890.17  &11527.64$^c$& 95.3$\pm$3.2&  315$\pm$11\\
\ion{[Ne}{iii]}&  3968.53  &11761.36& 64.3$\pm$3.4&  209$\pm$11\\
H7             &  3971.20  &11767.48& 51.8$\pm$1.7&  168$\pm$6\\
\hdelta        &  4102.92  &12158.43& 128.5$\pm$2.7& 403$\pm$9\\
\hgamma        &  4341.69  &12866.19& 205.9$\pm$1.3& 610$\pm$4\\
\hbeta         &  4862.70  &14410.72$^c$& 500.0$\pm$0.9& 1324$\pm$2\\
\ion{[O}{iii]} &  4960.29  &14700.41$^c$& 669.5$\pm$1.9& 1739$\pm$5\\
\ion{[O}{iii]} &  5008.24  &14841.33$^c$& 2357$\pm$3.5&  6066$\pm$9\\    
\ion{He}{i}    &  5877.28  &17417.44& 47.0$\pm$5.2&  104$\pm$12\\ 
\halpha        &  6564.63  &19453.85$^c$& 1802.1$\pm$0.6& 3645$\pm$1\\
\ion{[N}{ii]}  &  6585.42  &19513.96$^c$& 58.1$\pm$2.7 & 117$\pm$5\\
\ion{[S}{ii]}  &  6718.29  &19910.44& 156.8$\pm$1.1& 311$\pm$2\\
\ion{[S}{ii]}  &  6732.67  &19951.76& 102.7$\pm$1.7& 203$\pm$3\\

\hline        
\end{tabular}
\label{tab:m0304}

\textit{Notes.~}\\ 
$^a$ Rest-frame vacuum wavelengths.\\
$^b$ Emission line flux in units of $10^{-18}$ \ecs. $f_{\mathrm{obs}}$ and
$f_{\mathrm{cor}}$ correspond to the observed flux and that
corrected for intrinsic absorption, respectively.\\
$^c$ Affected by telluric absorption lines.\\
\end{table}


\subsection{SMACS\,J2031.8--4036 ID 1.1}

Emission lines from this arc are listed in Table~\ref{tab:m2031}. The
detected lines, excluding \lya, give a flux-weighted average redshift
of $z=3.5073\pm0.0002$.  At the low S/N per pixel in the UVB and VIS
arms, we can only identify several Lyman forest absorption lines, and
\ion{Si}{ii}\,$\lambda$1206 and \ion{S}{iv} $\lambda\lambda$1394,1403
at $z=3.5061\pm0.0013$. A detailed analysis of the absorption lines
requires a spectrum with a higher S/N measured in the continuum, so we
refrain from any detailed analysis of the present data.  Again, the
\lya\ emission line profile is double-peaked, with a rest-frame EW of
20~{\AA} (see Fig.~\ref{fig:Lyafit}), and we will return in
Sect. ~\ref{sect:model_lya} with a more detailed treatment of this
line.

At a low S/N we detect the semi-forbidden \ion{N}{iv]} $\lambda$1486
  line which is rarely seen in high-$z$ galaxies, although a few
  detections of this line in emission are reported for lensed
  galaxies, for example cB58 \citep{pettini00}, the Lynx arc at
  $z=3.3$ \citep{fosbury03}, and a galaxy at $z=5.563$
  \citep{raiter10}. When this line is detected, it can be interpreted
  as being caused by a large population of young and very hot, massive
  stars, but also an obscured AGN could be responsible for the
  ionisation \citep{vanzella10a}. Since neither \ion{N}{v}
  $\lambda\lambda$1238,1242 nor \ion{He}{ii} $\lambda$1640 are
  detected, we exclude the possibility that the ionisation is caused
  by an AGN \citep[see Table 1 in][]{binette03}. Hidden AGN like in
  radio galaxies typically display strong \ion{He}{ii} $\lambda$1640
  lines \citep[e.g.][]{villar-martin99}.

\begin{table}
    \caption{Emission lines from the SMACS J2031 arc}
  \begin{tabular}{lccrr}
\hline
\hline
   line       &  $\lambda_r$ $^a$   & $\lambda_{\mathrm{obs}}$ &
   $f_{\mathrm{obs}}^b$ & $f_{\mathrm{cor}}^b$ \\
\hline
\lya           &   1215.67  & 5480.80 &173.7$\pm$0.5 & 242$\pm$1 \\
\ion{N}{iv]}   &   1486.50  & 6700.85  & 7.6$\pm$0.9 & 10.1$\pm$1.2 \\
\ion{O}{iii]}  &   1660.81  & 7485.15  & 2.9$\pm$0.6 & 3.8$\pm$0.8\\
\ion{O}{iii]}  &   1666.15  & 7509.46  & 8.8$\pm$0.7 & 11.5$\pm$0.8\\
\ion{[C}{iii]} &   1906.68  & 8594.04  &12.0$\pm$0.5 & 15.4$\pm$0.6\\
\ion{C}{iii]}  &   1908.73  & 8603.54  & 8.4$\pm$0.9 & 10.8$\pm$1.2\\
\ion{[O}{ii]}  &   3727.09  & 16800.18 &14.4$\pm$1.6 & 16.9$\pm$1.9\\
\ion{[O}{ii]}  &   3729.88  & 16811.01 &15.5$\pm$1.3 & 18.2$\pm$1.5\\
\ion{[Ne}{iii]}&   3869.84  & 17442.96 &14.8$\pm$2.1 & 17.3$\pm$2.5 \\
\ion{H}{i} (H7)&   3971.20  & 17898.76 & 6.0$\pm$1.0 & 7.0$\pm$1.2\\ 
\hbeta         &   4862.70  & 21917.47 &37.7$\pm$1.3 & 42.8$\pm$1.5\\ 
\ion{[O}{iii]} &   4960.29  & 22357.68 &61.9$\pm$0.9 & 70.1$\pm$1.0\\      
\ion{[O}{iii]} &   5008.24  & 22573.68 &205.2$\pm$0.5& 232$\pm$0.6 \\
\hline        
\end{tabular}
\label{tab:m2031}

\textit{Notes.~} \\
$^a$ Rest-frame vacuum wavelengths.\\
$^b$ Emission line flux in units of $10^{-18}$ \ecs. $f_{\mathrm{obs}}$ and
$f_{\mathrm{cor}}$ correspond to the observed flux and that
corrected for intrinsic absorption, respectively.\\
\end{table}


\section{Physical conditions from emission lines}
\label{sect:emphys}
All emission line fluxes listed in Tables 1--3 are corrected for
Galactic extinction using the dust maps in \citet{schlegel98} along
with an extinction curve in \citet{fitzpatrick99} with $R_V$=3.1.  All
the physical properties derived in this section are summarised in
Table~\ref{tab:phys_prop}. When line ratios are considered, we do not
apply any correction for slit losses and magnification factors, since
we assume that the emission lines come from the exact same region.

\subsection{Gas phase reddening}
\label{sect:ebv_gas}
Any ratio of Balmer lines can be used to derive the reddening in the
gas phase of the galaxies.  In the absence of reddening, the expected
emission-line ratio is a function of the gas temperature and
density. We adopt the values tabulated for temperatures
$T$~=~10~000--20~000 K and densities $n\sim100$ cm$^{-3}$ in
\citet{brocklehurst71}. The choices of these values are based on the
estimated gas temperature and density in Sect~\ref{sect:temden}.

To derive the gas phase reddening, $E(B-V)_{\mathrm{gas}}$, we use 
all of the Balmer line pairs available and calculate the weighted
average, such that the reddening is determined by the Balmer line
ratio with the smallest uncertainty. We assume an extinction curve
derived for starburst galaxies \citep{calzetti00}, and the resulting
reddening values are listed in Table~\ref{tab:phys_prop}. The
reddening determined for M2031 is uncertain, and a value of zero is
consistent with that determined from fitting the entire spectrum with
stellar population models as demonstrated in Paper I.

\subsection{Starburst/AGN ionisation}
The ionising radiation from either AGN or recent formation of massive
stars gives rise to strong emission lines in galaxies. Whether the
narrow emission lines are caused by an AGN rather than a strong 
starburst is usually investigated through emission-line ratios, which
are sensitive to the hard ionisation from AGN. When only the strongest
rest-frame optical emission lines can be detected, as is typically the
case in high-redshift galaxies, the line ratios
\ion{[O}{iii]}/\hbeta\, versus \ion{[N}{ii]}/\halpha\ can be used to
distinguish the dominant ionising source \citep{kewley02}. Among the
galaxies anaysed in this paper, only M0304 have all the relevant lines
detected, and its line ratios are characteristic of star-forming
galaxies.

In M2031, the \ion{C}{iv}\,$\lambda\lambda$1548,1550 lines are not
detected to a level of $3\times10^{-18}$ \ecs, so the ratio
\ion{C}{iv} $\lambda\lambda$1548,1550/[\ion{C}{iii}]\ion{C}{iii]
$\lambda\lambda$1907,1909}$<$0.15 indicates a softer ionising
  spectrum, while an AGN would cause a ratio of 2 as argued for the
  Lynx arc \citep{binette03}. The line ratio for A31.1 is
  \ion{C}{iv}/\ion{C}{iii]}~=~0.65$\pm$0.71 indicating a higher
    ionisation parameter. In AGNs, the lines \ion{N}{v}
    $\lambda\lambda$1238,1242 and \ion{He}{ii} $\lambda$1640 are
    expected to be strong, and since these lines are not detected in
    any of our spectra it suggests that the main contributor to the
    ionisation is massive stars.

\subsection{Gas temperatures and densities}
\label{sect:temden}

Although \ion{[O}{iii]}\,$\lambda$4363, which is conventionally used
for direct temperature measurements in low-redshift galaxies, is not
detected in the M2031 source, the electron temperature can instead be
derived directly from the
\ion{O}{iii]}\,$\lambda\lambda$1661,1666/\ion{[O}{iii]}\,$\lambda$5007
  ratio as demonstrated for the Lynx arc \citep{villar-martin04}.  We
  use the {\tt nebular} package in IRAF \citep{shaw95} to determine
  the gas phase temperatures and densities. From the oxygen line-flux
  ratio, we find $T_e=16100\pm300$~K.  The
  \ion{[C}{iii]},\ion{C}{iii]}\,$\lambda\lambda$1907,1909 doublet can
    be used to derive densities. For M2031 we determine
    $n_e=2900\pm5000$ cm$^{-3}$ given that $T_e=16100$ K. For the
    M0304 source, we also use the
    \ion{O}{iii]}\,$\lambda\lambda$1661,1666 /
      [\ion{O}{iii}]\,$\lambda$5007 ratio to derive the temperature
      and density directly, again after correcting the measured
      fluxes for reddening. The results are listed in
      Table~\ref{tab:phys_prop}.

Two different diagnostics can be used to derive the electron
temperature for the Abell 1689 31.1 source. Using the
\ion{O}{iii]}\,$\lambda\lambda$1661,1666 lines gives
  $T_e=26900\pm1000$ K, while the commonly used \ion{[O}{iii]} 4363
  gives $T_e=21600\pm3000$ K as determined iteratively using the
  equations in \citet{aller84}.  Since the ratio from \ion{[O}{iii]}
  $\lambda4363$ over [\ion{O}{iii}] $\lambda\lambda$4959,5007 is less
  affected by reddening, we chose to use the latter temperature
  measurement for further analysis of the oxygen abundance. The ratio
  of the \ion{[C}{iii]},\ion{C}{iii]}\,$\lambda\lambda$1907,1909
    ($1.9\pm0.2$) doublet is unphysical at the 1.2$\sigma$ level, as
    the maximum allowed fraction is 1.65, however, the high value is
    consistent with a low electron density.

 Since the [\ion{O}{ii}] $\lambda\lambda$3727,3730 doublet is detected
 in all three galaxies, we can also use this to determine the electron
 density as listed in Table~\ref{tab:phys_prop}. The accurately
 measured electron densities from the medium resolution data of the
 three galaxies is similar to the densities in other lensed $z\sim2$
 galaxies where a range between a few 100 and a few 1000 cm$^{-3}$ are
 common \citep{hainline09,wuyts12}.

\subsection{Abundances}
In this section we derive the integrated abundances of oxygen and
carbon. These values are not dependent on magnification, because these
parameters are derived from line-flux ratios rather than absolute
values.

\subsubsection{Direct ($T_e$) abundance measurements}
\label{sect:direct}
Since we have measured the temperature directly for the sources, we
can determine the oxygen abundance with direct temperature methods. We
use the equations in \citet{izotov06} with the atomic data referenced
in \citet{stasinska05} to derive O$^{2+}$/H$^+$ and O$^{+}$/H$^+$. The
sum of these two contributions gives the total oxygen abundance, when
neglecting contributions from higher ionised stages, as is
conventionally done. We take into account that the two different ions
may reside in different zones of the \ion{H}{ii} region, and therefore
have different electron temperatures. Since we do not detect the
[\ion{O}{ii}] $\lambda\lambda$7320,7331 lines in any of the spectra,
we assume a relation between the temperatures $T$(\ion{O}{ii}) and
$T$(\ion{O}{iii}) described by equation 14 in \citet{izotov06}. The
resulting oxygen abundances are listed in Table~\ref{tab:phys_prop}.

Using the temperature and densities determined in the section above,
we calculate the total oxygen abundance for M2031.  The derived
uncertainty includes both the uncertainty from the emission line
fluxes, temperature, and density.  The contribution from singly
ionised oxygen is 12+log(O$^+$/H$^+$)~=~$6.94\pm0.06$ and double
ionised oxygen is 12+log(O$^{2+}$/H$^+$)~=~$7.69\pm0.02$. The sum
gives the total oxygen abundance 12+log(O/H)~=~$7.76\pm0.03$, assuming
that other oxygen ionisations states do not contribute.

For M0304 the contribution from single and double-ionised oxygen is
12+log(O$^+$/H$^+$)~=~$7.61\pm0.11$ and
12+log(O$^{2+}$/H$^+$)~=~$7.89\pm0.08$, respectively. The sum gives
the total oxygen abundance 12+log(O/H)~=~$8.07\pm0.09$, again assuming
that other oxygen ionisation states do not contribute.

For Abell 1689 ID 31.1 the contribution from single and double-ionised
oxygen is 12+log(O$^+$/H$^+$)~=~$6.42\pm0.33$ and
12+log(O$^{2+}$/H$^+$)~=~$7.66\pm0.11$, respectively, and the total
oxygen abundance is 12+log(O/H)~=~$7.69\pm0.13$ with the
$T$(\ion{O}{iii}) temperature determined from [\ion{O}{iii]}
$\lambda$4363 in Sect.~\ref{sect:temden}.

\subsubsection{Strong-line diagnostics}
\label{sect:R23}
In high-redshift galaxies temperature-sensitive lines are typically
not detected. Instead one has to rely on the relations between strong
emission line ratios and the directly derived oxygen abundances. The
ratio defined as \( R_{23} = (\mathrm{[OII]}\lambda\lambda3727,3730 +
\mathrm{[OIII]}\lambda\lambda4959,5007)/\hbeta\) introduced by
\citet{pagel79} is widely used, and its calibration relies either on
photo-ionisation models
\citep[e.g.,][]{mcgaugh91,kobulnicky99,kewley02}, or comparison with
objects where direct oxygen abundances are derived from
temperature-sensitive lines
\citep[e.g.,][]{alloin79,pilyugin05,nagao06}.  The $R_{23}$ relation
has two possible solutions with a high-metallicity and a
low-metallicity branch. A degenerate solution can be avoided if other
emission lines, such as [\ion{N}{ii}] $\lambda$6586 or [\ion{Ne}{iii}]
$\lambda$3869, are detected
\citep{denicolo02,kewley02,pettini04,liang06}. Also the [\ion{O}{iii}]
$\lambda$5007/[\ion{O}{ii}] $\lambda$3727 line ratio can be used to
distinguish between the upper and lower branch of the $R_{23}$
calibrations \citep{nagao06}. Any strong emission line diagnostics
have large intrinsic scatters of $\sim$0.2 dex, while various
calibrations may differ by up to 0.7 dex relative to each other
\citep{kewley08}, thereby making it difficult to compare objects from
the literature.

The determinations of oxygen abundances using strong-line diagnostics
were presented in Paper I, where we used [\ion{O}{ii}],[\ion{O}{iii}],
or [\ion{Ne}{iii}] emission line ratios to distinguish between the
upper and lower branch of the $R_{23}$ calibration using the method in
\citet{pilyugin05}, the \emph{O3N2} in \citet{pettini04} and \emph{Ne3O2}
in \citet{nagao06}.  For direct comparison, the derived oxygen
abundances are reproduced in Table~\ref{tab:phys_prop}. In addition to
the uncertainty of the abundance derived from the emission line
ratios, the calibrations themselves have an intrinsic scatter of
$\sim$0.2 dex, which must be taken into account when comparing the
strong-line diagnostics with the direct measurements. In any case, it
is encouraging that the strong-line calibration, in particular the one
in \citet{pilyugin05} agrees very well with the direct oxygen
abundances.

\subsubsection{C/O abundance}
\label{sect:co}
The abundance ratio C/O increases with increasing oxygen abundance
above 10\% solar. This trend is detected in halo stars, \ion{H}{ii}
regions \citep{garnett95}, and in the integrated spectra of Lyman
break galaxies \citep{shapley03}. A sub-solar C/O ratio arises when the
production of O and C is dominated by SN Type II explosions, while C
can also be produced by intermediate mass stars.  Another explanation
for the variation of the C/O ratio as a function of the oxygen
abundance involves a metallicity dependence of the mass-loss rate for
high-mass stars \citep{akerman04}.

Following the method in \citet{erb10} we use the ratio of
\ion{O}{iii]} $\lambda\lambda$1661,1666 and
  \ion{[C}{iii]},\ion{C}{iii}] $\lambda\lambda$1907,1909 emission
    lines to determine the C/O abundance ratios. Since the ionisation
    potentials for oxygen is higher than for carbon, C$^{3+}$ could be
    present in a medium with a high ionisation parameter.  The
    ionisation parameter $U$ is defined as the ratio between the
    density of ionising photons and the density of hydrogen atoms.

In the spectrum of M0304, the \ion{[C}{iii]},\ion{C}{iii}]
  $\lambda\lambda$1907,1909 doublet is not detected, even though it
  could have been, if the lines were as bright or brighter than
  \ion{O}{iii]} $\lambda\lambda$1661,1666. This is consistent with the
    note by \citet{shapley03} that the \ion{[C}{iii]},\ion{C}{iii}]
      lines appear to be stronger in LBGs with \lya\ in emission.

Once we know the metallicity of a galaxy, the ratio of the emission
lines [\ion{O}{iii}] $\lambda\lambda$4959,5007/[\ion{O}{ii}]
$\lambda\lambda$3727,3730 can be used to determine the ionisation
parameter using the photoionisation models and iterative equations in
\citet{kewley02}.  The three galaxies all have log$U$ between --3 and
--2 as listed in Table~\ref{tab:phys_prop}. Similar high ionisation
parameters have been measured for other gravitationally lensed
galaxies \citep{hainline09,richard11,rigby11}, as well as changes in
other emission lines ratios in LBGs relative to local galaxies have
been attributed to higher ionisation parameters \citep{erb06c,erb10}.

For an ionisation parameter of log~$U\approx-2$, the dominant
ionisation level for both C and O are double ionised, and one can
approximate C/O$\sim$C$^{2+}$/O$^{2+}$ \citep{erb10}, while at higher
ionisation parameters, an ionisation correction factor (ICF) is
necessary to include.  Given the values of the ionisation parameters
in the three lensed galaxies, no ICFs are needed to derive the C/O
ratio.  For M2031 we calculate
log(C$^{2+}$/O$^{2+}$)~=~$-0.80\pm0.09$, while for Abell 1869 ID 31.1
we derive log(C$^{2+}$/O$^{2+}$)~=~$-1.03\pm0.08$ based on the same
emission line ratio diagnostics. The results are listed in
Table~\ref{tab:phys_prop}. Compared to the solar value
log(C/O)$_{\odot}$~=~--0.26 \citep{asplund09}, the observed ratios are
sub-solar.  A decreasing C/O ratio with decreasing metallicity is a
known property of un-lensed LBGs \citep{shapley03,erb10}, and the low
C/O ratios of the lensed galaxies measured here are consistent with
their $\sim$0.1 solar oxygen abundance.

The fact that we observe a low C/O ratio when the galaxy ages are very
young ($\sim$4 Myr according to SED models in Paper I), suggests that
intermediate-mass stars have not contributed to the ISM in the
galaxies. As concluded by \citet{erb10} a C/O value consistent with
that observed in local metal-poor \ion{H}{ii} regions suggests that
the carbon production in high-mass stars depends on their
metallicities as predicted by either nucleosynthesis models
\citep{weaver93} or by mass loss from stellar winds \citep{maeder92}.

\subsubsection{N/O abundance}
\label{sect:no}
The nitrogen to oxygen ratio is an interesting parameter, because
nitrogen provides a measure of the chemical enrichment history of a
galaxy \citep[see][]{petitjean08}. Nitrogen can be produces in two
ways: a primary production from newly synthesised carbon via the CNO
cycle, and a secondary production also from a CNO cycle, but with
enhancement from a previous generation of stars that create additional
carbon and oxygen.

To determine the N/O abundance ratio for M0304, which is the only
source in the sample where [\ion{N}{ii}] $\lambda$6586 is detected, we
use the calibration of log(N$^+$/O$^+$) in \citet{pagel92}. To account
for the different ionisation stages arising in different zones in the
\ion{H}{ii} region, we convert the temperature measured for
$T$(\ion{O}{iii}) to $T$(\ion{O}{ii}) as in \citet{izotov06}.  As
argued by \citet{thurston96}, ionisation models of low-metallicity gas
show that log(N/O)~$\approx$~log(N$^+$/O$^+$), and that an ionisation
correction factor is not necessary. For the M0304 source we find
log(N/O)~=~~$-1.64\pm0.05$, which is below the plateau seen in
low-metallicity \ion{H}{ii} regions in dwarf and irregular galaxies
\citep{vanzee06}. The observed (N/O) ratio is more similar to the
ratios observed in the more metal-rich ($\sim$0.1 solar) damped Lyman-
$\alpha$ systems \citep{petitjean08,pettini08b,battisti12}, and
considerably below the solar level of log(N/O)$_{\odot}=-0.86$
\citep{asplund09}. This sub-solar (N/O) ratio suggests that the
galaxy, like in high-redshift damped Lyman-$\alpha$ absorbers, has
only experienced the primary production of nitrogen from the CNO
cycle, and cannot have had a large population of stars forming earlier
because these would have increased the nitrogen level. This is
consistent with a young age of $\sim$60 Myr based on stellar
population model fits in Paper I.

\begin{table*}
\begin{center}
\caption{Physical properties of sources with emission lines}
\begin{tabular}{lllll}
\hline
\hline
Arc         & A31.1  &  M0304  & M2031  & Section\\
\hline
Magnification $\mu$  & $26.6\pm3.1$ & $42.0\pm8.0$  & $15.8\pm7.0$ & Paper I\\

$z_{\mathrm{em}}-z_{\mathrm{abs}}$ (km s$^{-1}$) & $95\pm131$ &
$-10\pm55$ & $80\pm90$  & Paper I\\

\ebv$_{\mathrm{gas}}$ (mag) & $0.37\pm0.28$ & $0.23\pm0.01$ &
$0.03\pm0.33$ & \ref{sect:ebv_gas} \\

$f_{\mathrm{esc,LL}}$    &   ---     &   ---    &  $<$11\% & \ref{sect:fesc}\\
$f_{\mathrm{esc,\lya}}$  &  85\%  & 0\% & $20-50$\% & \ref{sect:model_lya}\\

\hline
$T_e$ (K)  & $21600\pm3000$  & $12900\pm900$ & $16100\pm300$  &
\ref{sect:temden} \\ 
$n_e$ \ion{[O}{ii]} (cm$^{-3}$) & $144\pm122$ & $81\pm8$ & $330\pm186$
& \ref{sect:temden} \\
$n_e$ \ion{[C}{iii]} (cm$^{-3}$) & ---& ---&$2900\pm5000$ &
\ref{sect:temden} \\

\hline

\smallskip
log $R_{23}$ & $0.85\pm0.05$ & $0.94\pm0.001$ & $0.91\pm0.02$ &
\ref{sect:R23} \\
12+log(O/H) ($R_{23}$) &  $7.63\pm0.10$ &   $8.18\pm0.05$ & $7.74\pm0.03$ & Paper I\\

\smallskip

$Z/Z_{\odot,R_{23}}$ & 0.1 & 0.3 & 0.1 & Paper I\\

12+log(O/H) (\emph{O3N2}) &  ---          & $8.04\pm0.01$ & ---        & Paper I\\
\smallskip

12+log(O/H) (\emph{Ne3O2})& $7.85\pm0.15$ & $8.20\pm0.01$ &
$7.56\pm0.11$ & Paper I\\

12+log(O/H) (direct)  &   $7.69\pm0.13$  & $8.07\pm0.09$ &  $7.76\pm0.03$  & \ref{sect:direct} \\ 
$Z/Z_{\odot,\mathrm{direct}}$  &   0.07  & 0.2 &   0.1 &   \\

log$U$       & $-2.1$   & $-2.8$ & $-2.1$     & \ref{sect:co} \\
log(C/O)  & $-1.03\pm0.08$  & --- & $-0.80\pm0.09$ & \ref{sect:co} \\
log(N/O)  &       ---       & $-1.64\pm0.05$ & ---& \ref{sect:no}\\
\hline
\end{tabular}
\label{tab:phys_prop}
\end{center}

\end{table*}

\subsection{Lyman-$\alpha$ escape fraction}
\label{sect:fesc_lya}
Two sources, M2031 and A1689 ID 31.1 have double-peaked \lya\ emission
lines, which is not uncommon as it is seen in 30\% of LBGs at $z=2-3$
\citep{kulas12}, and up to 50\% of Lyman-$\alpha$ emitters
\citep{yamada12}. Only few lensed galaxies studied in detail to date
are also strong \lya\ emitters \citep{fosbury03,quider09,bayliss10},
so the present data allow for a more in depth analysis including
constraints from the rest-frame optical emission lines. The escape
fraction can be determined from the observed Balmer emission lines
relative to the observed \lya\ line flux.  Under the assumption of a
case B recombination scenario, and zero extinction, we find that the
escape fraction of \lya\ photons from M2031 is 20\%, and 4\% or 85\%
for the A1689 31.1 arc before and after correction for an intrinsic
reddening of $E(B-V)=0.37$.

At comparable redshifts, the comparison of galaxies UV luminosity to
their \lya\ emission lines have revealed a median \lya\ escape
fraction of 29\%, where any value between 0 and 100 \% is possible
\citep{blanc11}. Similarly to our calculations, comparisons of the
flux from \halpha\ and \lya\ lines for both $z\sim0.3$ galaxies
\citep{atek09} and a few $z\sim2$ galaxies \citep{hayes10} indicate a
large range of \lya\ escape fractions from 0 to 100\%. There is a
tendency that larger escape fractions are found for galaxies with
smaller reddenings \citep{hayes10,blanc11}, but only when the
\lya\ emission lines are not corrected for the intrinsic reddening,
similar to what we find for the two \lya\ emitters in this study.

\subsection{Lyman limit escape fraction}
\label{sect:fesc}
The M2031 source has a sufficiently high redshift that we can derive
an upper limit on the escape fraction bluewards of the Lyman limit. In
the region $3800-4100$~{\AA}, corresponding to a rest-frame wavelength
of $\sim900$~{\AA}, the average flux measured in the extracted 1D
spectrum after binning by a factor of 50 in the dispersion is
$(1\pm5)\times10^{-20}$ \ecsa, while the observed flux at the
rest-frame 1500~{\AA} is $(2.3\pm0.3)\times10^{-18}$ \ecsa. The
relative escape fraction at these wavelengths needs to be corrected
for the absorption in the intergalactic medium (IGM)
\citep[e.g.,][]{inoue06,boutsia11}:
\[f_{\mathrm{esc,LL}}=
  \frac{(F_{1500}/F_{900})_{\mathrm{int}}}{(f_{1500}/f_{900})_{\mathrm{obs}}}
  \exp(-\tau_{900}^{\mathrm{IGM}}), \] where
  \(T=\exp(-\tau_{900}^{\mathrm{IGM}})\) is the transmission in the
  IGM at the rest-frame 900 {\AA}.  The intrinsic fraction of the flux
  densities $(F_{1500}/F_{900})_{\mathrm{int}}$ is model dependent,
  and we determine the ratio from the best fit spectral template in
  Paper I to be 1.44. Since we do not find evidence for significant
  intrinsic reddening, this value is not corrected for extinction.
  From the observed ratio and an optical depth of the IGM at $z=3.5$
  due to the Lyman series \citep{madau95} giving a transmission of 0.2
  at \(\lambda=900(1+3.5)\) {\AA}, we derive an upper limit for the
  escape fraction
$f_{\mathrm{esc,LL}} < 0.031\pm0.025$, or $<$3.0$\pm$2.5\%. However, the
  IGM transmission varies significantly along different sight lines
  due to the random occurrence of intervening Lyman limit systems, and
  at $z\sim3.5$ the IGM transmission just bluewards of the Lyman limit
  is $T=0.3\pm0.2$ based on numerical simulations \citep{inoue08}.
  Accordingly, our limit to the escape fraction is less constrained:
  $f_{\mathrm{esc,LL}}<$11\%.

In a sample of 14 LBGs at $z\sim3$ \citet{shapley06} found a
significant escape fraction in two galaxies implying an average escape
fraction of 14\%. However, because spatial offsets between the galaxy
emission in the rest-frame Lyman continuum and at rest-frame optical
wavelenghts have been measured, the large escape fraction could be
contaminated by lower redshift interlopers. Consequently,
\citet{vanzella10b} find that LBGs at $z\sim4$ have upper limit of
$f_{\mathrm{esc,LL}}<$~5--20\%. A stacked spectrum of 11 LBGs at
$z=3.3$ gave an upper limit of $<$~5\% \citep{boutsia11}, albeit
dominated by a couple of bright LBGs, which may not be representative
for the average high-redshift galaxy population. \citet{shapley06} and
\citet{boutsia11} use Starburst99 templates \citep{leitherer99}, with
an intrinsic flux ratio of three between 1500 and 900 {\AA}. A larger
flux ratio is the result of an older stellar population, which in turn
depends on the assumed star formation history. If we assume the same
intrinsic ratio (i.e. 3), the corresponding limit on the escape
fraction derived for the M2031 galaxy would be a factor of two higher,
i.e., $f_{\mathrm{esc,LL}}<$22\%.

Even though lensing significantly boosts the signal from faint
galaxies such that we can detect very small fluxes at rest-frame
900~{\AA} for the M2031 source, which has an intrinsic UV luminosity
of 9$L^*$ determined from the SED fits in Paper I, we may have to look
for intrinsically much fainter galaxies ($<$0.1$L^*$ ) that are
expected to dominate the escaping UV flux at high redshifts.  Future
observations of gravitationally lensed faint galaxies offer an
opportunity for determining $f_{\mathrm{esc,LL}}$ \citep{vanzella12}.


\section{Modelling the \lya\ emission lines}
\label{sect:model_lya}
To infer the physical conditions governing the two systems for which
\lya\ emission is detected, we undertake a series radiative transfer
(RT) simulations, with the aim of fitting synthetic spectra to the
observed ones.  For this purpose, we apply the three-dimensional
\lya\ scattering code {\sc MoCaLaTA} \citep{laursen09a}, including the
effects of dust \citep{laursen09b}. In the following, the basics of
the simulations are outlined.

The galaxies are modelled as spherical conglomerations of gas existing
in two phases: warm, chiefly neutral clouds of relatively high neutral
hydrogen density, dispersed in a hotter and more ionised intercloud
medium (ICM).  The ICM density, as well as the number density of
clouds, decreases exponentially with distance from the
center. Additionally, the systems are surrounded by a spherical shell
of outflowing gas.

Densities, temperatures, and velocity fields are assigned to cells in
a Cartesian grid. The grid has a base resolution of $128^3$ cells,
with cells constituting the edge of clouds being refined recursively
into eight cells, to make clouds more spherical.  In addition,
\lya\ photons are also emitted from the centre and out with a
luminosity which decreases exponentially with radius. A photon is
traced as it scatters its way out of the galaxy on individual hydrogen
atoms, constantly changing direction and frequency, or, alternatively,
is absorbed by dust. To obtain good statistics for the spectrum, a
total of $\sim 10^5$ photons are emitted.

At the redshifts of the two emitters ($z = 1.8$ and $z = 3.5$ for
\A\ and M2031, respectively), the IGM is ionised to a high degree.
Nevertheless, a fraction of the photons blueward of the \lya\ line may
still be scattered out of the line of sight by diffuse neutral gas
clouds in the circumgalactic medium. This effect is modelled in a
statistical way following \citet{laursen11}.

In addition to matching the observed spectra, the preferred model
should also be consistent with the star formation rates (SFRs)
obtained in Paper I, as well as with the escape fractions inferred
from comparing the integrated \lya\ and \halpha\ fluxes. Under these
constraints, and considering the large number of free parameters in
the model, rather than searching the full parameter space as has been
done previously for a shell-only model of four parameters, including
gas and dust column density, temperature, and expansion velocity
\citep{verhamme08, schaerer11}, a series of simulations is conducted,
changing the various input parameters until a satisfactory fit is
obtained. This approach has been demonstrated to be able to match
simultaneously a broad range of observables for a single system
\citep{noterdaeme12}.

As an ``initial'' model, we make use of the few observational
constraints we have at hand. Since we have measured the metallicities
of the two systems (0.07 and 0.10 solar for \A\ and M2031,
respectively), this parameter is held fixed in the simulation to limit
the already high number of free parameters.

Assuming an SMC extinction law and a dust-to-metal ratio similar to
the local universe \citep[which is thought to be approximately valid
  even at high redshift; see, e.g.,][]{pei98, pettini97,savaglio03},
the dust density is calculated.  The {\sc MoCaLaTA} code can model the
dust as either LMC or SMC dust. We choose SMC dust, as the young
stellar population of the SMC may be expected to match more closely
those of the observed galaxies, but note that the difference between
various extinction laws is of minor importance; over the narrow
spectral region of the \lya\ line, the dust cross section is almost
flat, and indeed a comparison study in \citet{laursen09b} of the two
dust models reveals a change in escape fractions at the $\sim$1\%
level only.  Furthermore, the fact that the blue components of the
spectra are significantly smaller than the red ones, albeit still
visible, is a hint of modest outflow velocities.  Finally, for the
temperatures we assume typical values of the three phases in the
model: $10^4$, $2\times10^4$, and $10^5$ K for the clouds, the ICM,
and the expanding shell, respectively.

\subsection{Best fit models}
\begin{figure*}
\begin{center}
\includegraphics[width=.45\textwidth]{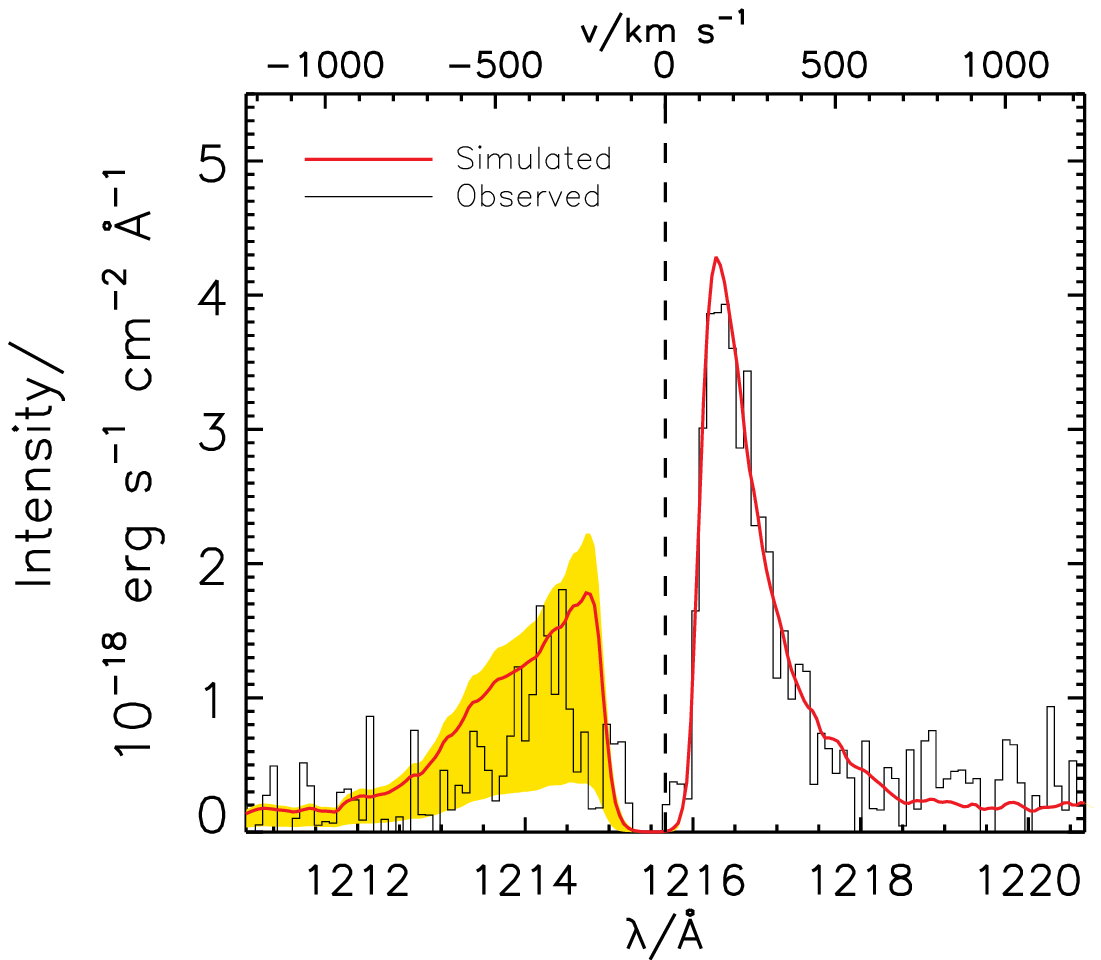}
\includegraphics[width=.45\textwidth]{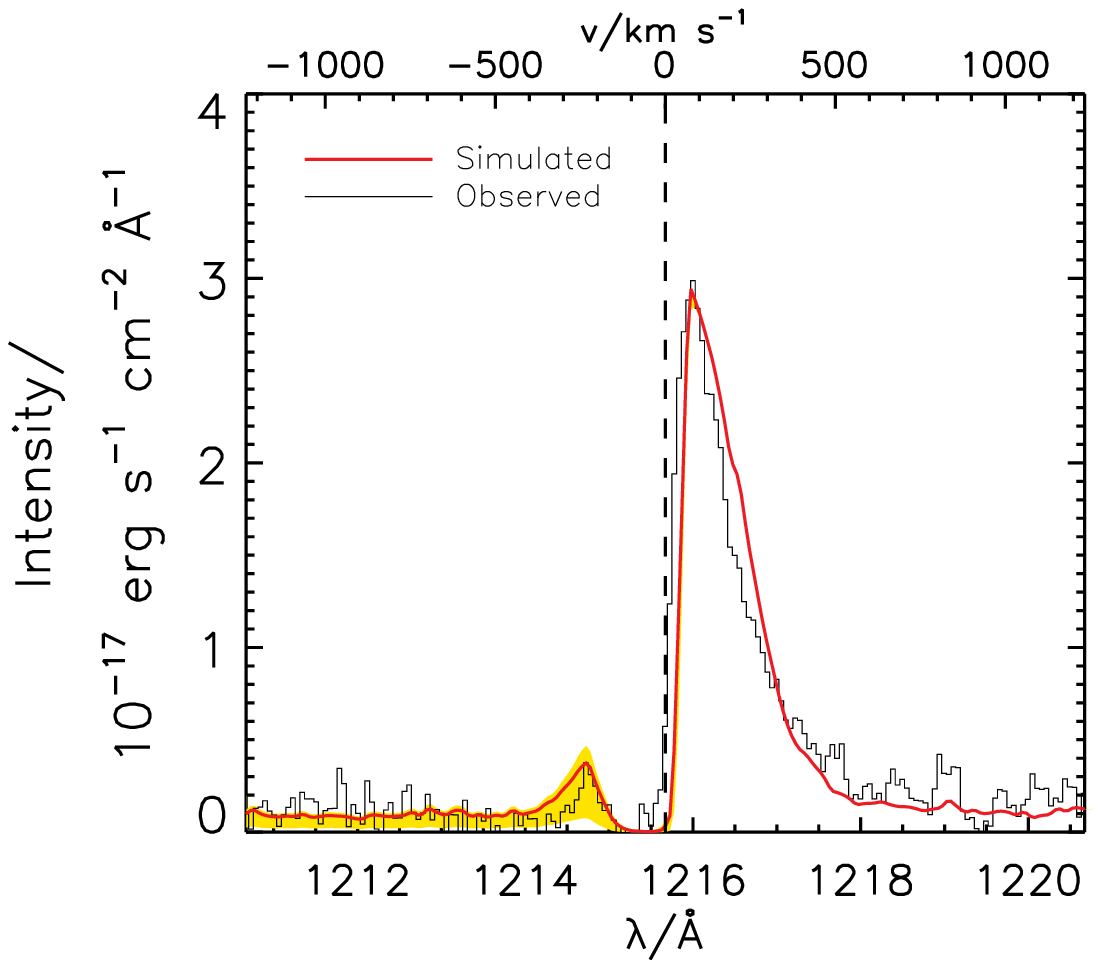}
\end{center}
\caption{Observed (\emph{black}) and fitted (\emph{red}) spectra of
  the systems \A ~(\emph{left}) and M2031 (\emph{right}). The
  \emph{yellow} region shows the 68\% confidence interval due to
  absorption in the IGM.}
\label{fig:Lyafit}
\end{figure*}

Figure \ref{fig:Lyafit} shows the best fit models to the observed
spectra, given the constraints. The synthetic spectra have been scaled
by an SFR of 0.8 and 4.5 $M_\odot$ yr$^{-1}$, respectively, which is
consistent with the values found in Sect.~\ref{sect:emphys}.
Moreover, the escape fractions were found to be 0.87 and 0.56,
respectively.  While the former matches perfectly the value found for
\A, the latter is somewhat higher than the $\sim$0.20 found for M2031,
although due to the rather large uncertainty on the fluxes, it is not
inconsistent.
Table \ref{tab:fitres} summarises the final parameter values used for
the models.
\begin{table}
\begin{center}
\caption{Galaxy model values}
\begin{tabular}{l|ll|}
\hline
\hline
Parameter                      & \A                           &  M2031    \\
\hline
Galaxy radius                  & 1 kpc                        & 1 kpc     \\
Number of clouds               & 80                           & 200       \\
Cloud radii                    & 20 pc                        & 20 pc     \\ 
Cloud temperature              & $10^4$ K                     & $10^4$ K  \\
Cloud density$^{a}$            & 2.5 cm$^{-3}$                & 100 cm$^{-3}$ \\
Metallicity                    & 0.07~$Z_{\odot}$             & 0.1~$Z_{\odot}$\\
Cloud velocity dispersion      & 50 \kms                      & 10 \kms   \\
Central ICM density$^{a}$      & 7.5$\times$$10^{-2}$ cm$^{-3}$ & 3$\times$$10^{-4}$ cm$^{-3}$ \\
ICM density$^{a}$ scale length & 0.4 kpc                      & 1 kpc     \\
ICM temperature                & $5\times10^4$ K              & $5\times10^4$ K \\
Shell radius                   & 1 kpc                        & 1 kpc     \\
Shell expansion velocity       & 60 \kms                      & 110 \kms  \\
Shell column density$^{a}$     & $3\times10^{18}$ cm$^{-2}$   & $10^{18}$ cm$^{-2}$ \\
Shell temperature              & $5\times10^5$ K              & $5\times10^5$ K \\
\lya\ emission scale length    & 0.1 kpc                      & 0.25 kpc  \\
Intrinsic \lya\ EW         & 40 {\AA}                     & 20 {\AA} \\
\hline
\end{tabular}
\label{tab:fitres}
\end{center}
\flushleft{$^{a}$All densities refer to \emph{neutral} hydrogen only.}
\end{table}

Note that the models are perfectly scalable: radii of 1 kpc and 2 kpc
have been assumed for \A\ and M2031, respectively; however, the exact
same spectra and escape fractions would emerge from models with twice
the radius, but with all densities a factor of 4 smaller, and scale
lengths and cloud radii doubled.


 As mentioned in Sec. \ref{sect:data}, the asymmetric profile is
 characteristic of an expanding ISM. However, the fact that we
 \emph{do} see the blue peaks in both spectra tells us that the
 typical outflow velocities must be quite low.  For a homogeneous
 medium, the blue peak vanishes already at $V_{\mathrm{out}} \sim 100$
 \kms\ for a broad range of column densities \citep[e.g.][]
           {verhamme06}. For an inhomogeneous medium, since the
           photons blueward of the line may find low-density paths out
           of the galaxy, higher velocities may be allowed with the
           blue peak still visible.

The outflows of the two models that match the observed spectra best
are 60 and 110 \kms\ for A31.1 and M2031, respectively. This is
consistent with the values inferred from comparing the redshifts of
the absorption and emission lines (see
Tab.~\ref{tab:phys_prop}. Although the rather large uncertainties in
the latter values make them consistent with zero as well, we consider
it likely that the two emitters in fact do exhibit outflow. In
principle, foreground diffuse \ion{H}{i} clouds could also cause a
decreased blue peak. At the given redshifts the IGM is likely to be
highly ionised. However, its clumpy nature could in principle cause
absorption nevertheless. In Fig.~\ref{fig:Lyafit}, the yellow region
marks the 68\% confidence level of the \lya\ emission line,
\emph{after} being transmitted through the IGM. Indeed, the expected
impact of the IGM is rather low.

Most observed \lya\ emitters exhibit an asymmetric line profile with
the blue peak missing. However, with sufficient spectral resolution it
seems that the fraction of \lya\ emitters where the blue peak is still
visible is quite large \citep [20--50\%][]{kulas12,yamada12}.

It is important to keep in mind that we do not claim that the model
galaxies must be a realistic representation of the real galaxies. Many
processes shape the \lya\ line, and these are to some extend
degenerate. Nevertheless, it is striking that it is possible to match
the lines under the restrictions set by the other observables such as
metallicity and SFR.

The higher SFR and metallicity of M2031 relative to A31.1
hints at the former being more evolved than the latter. This
interpretation is backed up by the results of the simulations; the
higher number of clouds and the larger density contrast between clouds
and ICM implies a more fragmented ISM. The shallower density gradient
could then be caused by the enhanced star formation ``inflating'' the
galaxy, while the higher wind velocity is a natural consequence of a
more vigorous feedback.


\section{Summary}
\label{sect:conclusions}

We have measured the oxygen abundance using direct $T_e$ methods for
three galaxies at $z$ = 1.8, 2.0 and 3.5, respectively. The direct
oxygen abundance measurements are in agreement with those derived from
the conventional $R_{23}$ ratio using the diagnostics calibration in
\citet{pilyugin05}. This proves that the strong-line diagnostics are
in fact useful to determine oxygen abundances in high-redshift
galaxies, as the temperature-sensitive lines for field galaxies are
almost always below the detection limit even on 8-m telescopes.

In addition, we have measured the C/O and N/O ratios for the three
galaxies, and argue that their ratios are consistent with the general
trend of sub-solar metal ratios for low-metallicity galaxies, and in
particular for young galaxies where contributions from either nitrogen
or carbon from older stars have not yet occured.

We have modeled the emission line profiles in two strong
\lya\ emitting galaxies using the constraints (SFR, reddening, escape
fractions and metallicity) obtained from the analysis of their
spectra. The relatively small inferred expansion velocities of the
neutral gas shells correspond to the small velocity offsets seen in
absorption and rest-frame emission lines. The observed absorption and
emission line redshifts show that the velocity offsets are consistent
with zero at the 1$\sigma$ confidence level, but are also consistent
with the small shell expansion velocities (60 and 110 km s$^{-1}$,
respectively) found in the radiative transfer simulations. Compared to
more massive LBGs with higher SFRs we observe a smaller velocity
offset between the absorption and \lya\ emission wavelengths. 
To develop our understanding how galaxy outflows are related to the
SFRs and escape of \lya\ photons, it would be interesting to expand
the sample and verify if the trend of decreasing velocity offset with
increasing \lya\ EW observed for LBGs \citep{shapley03} is present
also in \lya\ emitters with much higher EWs.

Recent studies of \lya\ emitters have shown that the absorption lines
are more difficult to detect at progressively larger \lya\ equivalent
widths, even when stacking the data from many emitters
\citep{berry12}. Gravitational lensing is the only means to obtain
medium to high-resolution spectroscopic data of high-EW \lya\ emitters
\citep{diehl09,quider09,bayliss10}. The A31.1 source, with an
equivalent width of 40 {\AA}, does not appear to have strong
absorption lines, but the faint continuum and low signal-to-noise
ratio of the spectrum prevents a more detailed analysis besides the
strongest \ion{Si}{ii} lines at 1260, 1264 and 1526 {\AA}. Since one
hour integration time per target was used in this study, we conclude
that it would be possible to investigate the detailed physics of a
strong \lya\ emitter with a few more hours integration time on a large
telescope.

Future observations of strong lensing clusters, such as the
\emph{HST}/CLASH survey \citep{postman12}, may reveal more potentially
interesting lensed high-redshift sources with steep UV slopes and high
surface-brightness regions. These could be the best targets to locate
and investigate the interstellar absorption lines in strong
\lya\ emitting galaxies. These galaxies are also the prime candidates
to find other strong UV lines such that abundances of several species
could be determined by direct methods. With such observations, we
could constrain better chemical evolution models of high-redshift
galaxies.

\section*{Acknowledgments}
The Dark Cosmology Centre is funded by the DNRF.  LC is supported by
the EU under a Marie Curie Intra-European Fellowship, contract
PIEF-GA-2010-274117.  PL acknowledges funding from the Villum
foundation. JR is supported by the Marie Curie Career Integration
Grant 294074. BMJ acknowledged support from the ERC-StG grant
EGGS-278202. We thank Stefano Covino, Valerio D'Elia, Johan
P.~U. Fynbo, Daniele Malesani, Hans Ulrik N{\o}rgaard-Nielsen, and
Beate Stelzer for carrying out the observations. The RT simulations
were performed on facilities provided by the Danish Center for
Scientific Computing.

\bibliographystyle{apj}
\bibliography{ms_lc2}

\begin{thebibliography}{}

\bibitem[\protect\citeauthoryear{{Akerman} et~al.}{{Akerman}
  et~al.}{2004}]{akerman04}
{Akerman}, C.~J., {Carigi}, L., {Nissen}, P.~E., {Pettini}, M.,  \& {Asplund},
  M. 2004, \aap, 414, 931

\bibitem[\protect\citeauthoryear{{Allam} et~al.}{{Allam}
  et~al.}{2007}]{allam07}
{Allam}, S.~S., {Tucker}, D.~L., {Lin}, H., {Diehl}, H.~T., {Annis}, J.,
  {Buckley-Geer}, E.~J.,  \& {Frieman}, J.~A. 2007, \apjl, 662, L51

\bibitem[\protect\citeauthoryear{{Aller}}{{Aller}}{1984}]{aller84}
{Aller}, L.~H., ed. 1984, Astrophysics and Space Science Library, Vol. 112,
  {Physics of thermal gaseous nebulae} ({Dordrecht: D. Reidel})

\bibitem[\protect\citeauthoryear{{Alloin} et~al.}{{Alloin}
  et~al.}{1979}]{alloin79}
{Alloin}, D., {Collin-Souffrin}, S., {Joly}, M.,  \& {Vigroux}, L. 1979, \aap,
  78, 200

\bibitem[\protect\citeauthoryear{{Asplund} et~al.}{{Asplund}
  et~al.}{2009}]{asplund09}
{Asplund}, M., {Grevesse}, N., {Sauval}, A.~J.,  \& {Scott}, P. 2009, \araa,
  47, 481

\bibitem[\protect\citeauthoryear{{Atek} et~al.}{{Atek} et~al.}{2009}]{atek09}
{Atek}, H., {Kunth}, D., {Schaerer}, D., {Hayes}, M., {Deharveng}, J.~M.,
  {{\"O}stlin}, G.,  \& {Mas-Hesse}, J.~M. 2009, \aap, 506, L1

\bibitem[\protect\citeauthoryear{{Battisti} et~al.}{{Battisti}
  et~al.}{2012}]{battisti12}
{Battisti}, A.~J., et~al. 2012, \apj, 744, 93

\bibitem[\protect\citeauthoryear{{Bayliss} et~al.}{{Bayliss}
  et~al.}{2010}]{bayliss10}
{Bayliss}, M.~B., {Wuyts}, E., {Sharon}, K., {Gladders}, M.~D., {Hennawi},
  J.~F., {Koester}, B.~P.,  \& {Dahle}, H. 2010, \apj, 720, 1559

\bibitem[\protect\citeauthoryear{{Berry} et~al.}{{Berry}
  et~al.}{2012}]{berry12}
{Berry}, M., et~al. 2012, \apj, 749, 4

\bibitem[\protect\citeauthoryear{{Bian} et~al.}{{Bian} et~al.}{2010}]{bian10}
{Bian}, F., et~al. 2010, \apj, 725, 1877

\bibitem[\protect\citeauthoryear{{Binette} et~al.}{{Binette}
  et~al.}{2003}]{binette03}
{Binette}, L., {Groves}, B., {Villar-Mart{\'{\i}}n}, M., {Fosbury}, R.~A.~E.,
  \& {Axon}, D.~J. 2003, \aap, 405, 975

\bibitem[\protect\citeauthoryear{{Blanc} et~al.}{{Blanc}
  et~al.}{2011}]{blanc11}
{Blanc}, G.~A., et~al. 2011, \apj, 736, 31

\bibitem[\protect\citeauthoryear{{Boutsia} et~al.}{{Boutsia}
  et~al.}{2011}]{boutsia11}
{Boutsia}, K., et~al. 2011, \apj, 736, 41

\bibitem[\protect\citeauthoryear{{Brocklehurst}}{{Brocklehurst}}{1971}]{brocklehurst71}
{Brocklehurst}, M. 1971, \mnras, 153, 471

\bibitem[\protect\citeauthoryear{{Cabanac} et~al.}{{Cabanac}
  et~al.}{2005}]{cabanac05}
{Cabanac}, R.~A., {Valls-Gabaud}, D., {Jaunsen}, A.~O., {Lidman}, C.,  \&
  {Jerjen}, H. 2005, \aap, 436, L21

\bibitem[\protect\citeauthoryear{{Calzetti} et~al.}{{Calzetti}
  et~al.}{2000}]{calzetti00}
{Calzetti}, D., {Armus}, L., {Bohlin}, R.~C., {Kinney}, A.~L., {Koornneef}, J.,
   \& {Storchi-Bergmann}, T. 2000, \apj, 533, 682

\bibitem[\protect\citeauthoryear{{Christensen} et~al.}{{Christensen}
  et~al.}{2010}]{christensen10}
{Christensen}, L., {D'Odorico}, S., {Pettini}, M., {Belokurov}, V., {Evans},
  N.~W., {Kellogg}, M.,  \& {Vernet}, J. 2010, \mnras, 406, 2616

\bibitem[\protect\citeauthoryear{{Christensen} et~al.}{{Christensen}
  et~al.}{2012}]{christensen12a}
{Christensen}, L., et~al. 2012, MNRAS accepted (paper I)

\bibitem[\protect\citeauthoryear{{Denicol{\'o}}, {Terlevich}, \&
  {Terlevich}}{{Denicol{\'o}} et~al.}{2002}]{denicolo02}
{Denicol{\'o}}, G., {Terlevich}, R.,  \& {Terlevich}, E. 2002, \mnras, 330, 69

\bibitem[\protect\citeauthoryear{{Dessauges-Zavadsky}
  et~al.}{{Dessauges-Zavadsky} et~al.}{2011}]{dessauges11}
{Dessauges-Zavadsky}, M., {Christensen}, L., {D'Odorico}, S., {Schaerer}, D.,
  \& {Richard}, J. 2011, \aap, 533, A15

\bibitem[\protect\citeauthoryear{{Dessauges-Zavadsky}
  et~al.}{{Dessauges-Zavadsky} et~al.}{2010}]{dessauges10}
{Dessauges-Zavadsky}, M., {D'Odorico}, S., {Schaerer}, D., {Modigliani}, A.,
  {Tapken}, C.,  \& {Vernet}, J. 2010, \aap, 510, A26

\bibitem[\protect\citeauthoryear{{Diehl} et~al.}{{Diehl}
  et~al.}{2009}]{diehl09}
{Diehl}, H.~T., et~al. 2009, \apj, 707, 686

\bibitem[\protect\citeauthoryear{{D'Odorico} et~al.}{{D'Odorico}
  et~al.}{2006}]{dodorico06}
{D'Odorico}, S., et~al. 2006, in Society of Photo-Optical Instrumentation
  Engineers (SPIE) Conference Series, Vol. 6269

\bibitem[\protect\citeauthoryear{{Ebeling}, {Edge}, \& {Henry}}{{Ebeling}
  et~al.}{2001}]{ebeling01}
{Ebeling}, H., {Edge}, A.~C.,  \& {Henry}, J.~P. 2001, \apj, 553, 668

\bibitem[\protect\citeauthoryear{{Erb} et~al.}{{Erb} et~al.}{2010}]{erb10}
{Erb}, D.~K., {Pettini}, M., {Shapley}, A.~E., {Steidel}, C.~C., {Law}, D.~R.,
  \& {Reddy}, N.~A. 2010, \apj, 719, 1168

\bibitem[\protect\citeauthoryear{{Erb} et~al.}{{Erb} et~al.}{2006}]{erb06c}
{Erb}, D.~K., {Shapley}, A.~E., {Pettini}, M., {Steidel}, C.~C., {Reddy},
  N.~A.,  \& {Adelberger}, K.~L. 2006, \apj, 644, 813

\bibitem[\protect\citeauthoryear{{Fitzpatrick}}{{Fitzpatrick}}{1999}]{fitzpatrick99}
{Fitzpatrick}, E.~L. 1999, \pasp, 111, 63

\bibitem[\protect\citeauthoryear{{Fosbury} et~al.}{{Fosbury}
  et~al.}{2003}]{fosbury03}
{Fosbury}, R.~A.~E., et~al. 2003, \apj, 596, 797

\bibitem[\protect\citeauthoryear{{Garnett} et~al.}{{Garnett}
  et~al.}{1995}]{garnett95}
{Garnett}, D.~R., {Skillman}, E.~D., {Dufour}, R.~J., {Peimbert}, M.,
  {Torres-Peimbert}, S., {Terlevich}, R., {Terlevich}, E.,  \& {Shields}, G.~A.
  1995, \apj, 443, 64

\bibitem[\protect\citeauthoryear{{Hainline} et~al.}{{Hainline}
  et~al.}{2009}]{hainline09}
{Hainline}, K.~N., {Shapley}, A.~E., {Kornei}, K.~A., {Pettini}, M.,
  {Buckley-Geer}, E., {Allam}, S.~S.,  \& {Tucker}, D.~L. 2009, \apj, 701, 52

\bibitem[\protect\citeauthoryear{{Hayes} et~al.}{{Hayes}
  et~al.}{2010}]{hayes10}
{Hayes}, M., et~al. 2010, \nat, 464, 562

\bibitem[\protect\citeauthoryear{{Holden} et~al.}{{Holden}
  et~al.}{2001}]{holden01}
{Holden}, B.~P., et~al. 2001, \aj, 122, 629

\bibitem[\protect\citeauthoryear{{Inoue} \& {Iwata}}{{Inoue} \&
  {Iwata}}{2008}]{inoue08}
{Inoue}, A.~K.,  \& {Iwata}, I. 2008, \mnras, 387, 1681

\bibitem[\protect\citeauthoryear{{Inoue}, {Iwata}, \& {Deharveng}}{{Inoue}
  et~al.}{2006}]{inoue06}
{Inoue}, A.~K., {Iwata}, I.,  \& {Deharveng}, J.-M. 2006, \mnras, 371, L1

\bibitem[\protect\citeauthoryear{{Izotov} et~al.}{{Izotov}
  et~al.}{2006}]{izotov06}
{Izotov}, Y.~I., {Stasi{\'n}ska}, G., {Meynet}, G., {Guseva}, N.~G.,  \&
  {Thuan}, T.~X. 2006, \aap, 448, 955

\bibitem[\protect\citeauthoryear{{Kewley} \& {Dopita}}{{Kewley} \&
  {Dopita}}{2002}]{kewley02}
{Kewley}, L.~J.,  \& {Dopita}, M.~A. 2002, \apjs, 142, 35

\bibitem[\protect\citeauthoryear{{Kewley} \& {Ellison}}{{Kewley} \&
  {Ellison}}{2008}]{kewley08}
{Kewley}, L.~J.,  \& {Ellison}, S.~L. 2008, \apj, 681, 1183

\bibitem[\protect\citeauthoryear{{Kobulnicky}, {Kennicutt}, \&
  {Pizagno}}{{Kobulnicky} et~al.}{1999}]{kobulnicky99}
{Kobulnicky}, H.~A., {Kennicutt}, R.~C., Jr.,  \& {Pizagno}, J.~L. 1999, \apj,
  514, 544

\bibitem[\protect\citeauthoryear{{Kulas} et~al.}{{Kulas}
  et~al.}{2012}]{kulas12}
{Kulas}, K.~R., {Shapley}, A.~E., {Kollmeier}, J.~A., {Zheng}, Z., {Steidel},
  C.~C.,  \& {Hainline}, K.~N. 2012, \apj, 745, 33

\bibitem[\protect\citeauthoryear{{Lara-L{\'o}pez} et~al.}{{Lara-L{\'o}pez}
  et~al.}{2010}]{lara-lopez10}
{Lara-L{\'o}pez}, M.~A., et~al. 2010, \aap, 521, L53

\bibitem[\protect\citeauthoryear{{Laursen}, {Razoumov}, \&
  {Sommer-Larsen}}{{Laursen} et~al.}{2009a}]{laursen09a}
{Laursen}, P., {Razoumov}, A.~O.,  \& {Sommer-Larsen}, J. 2009a, \apj, 696, 853

\bibitem[\protect\citeauthoryear{{Laursen}, {Sommer-Larsen}, \&
  {Andersen}}{{Laursen} et~al.}{2009b}]{laursen09b}
{Laursen}, P., {Sommer-Larsen}, J.,  \& {Andersen}, A.~C. 2009b, \apj, 704,
  1640

\bibitem[\protect\citeauthoryear{{Laursen}, {Sommer-Larsen}, \&
  {Razoumov}}{{Laursen} et~al.}{2011}]{laursen11}
{Laursen}, P., {Sommer-Larsen}, J.,  \& {Razoumov}, A.~O. 2011, \apj, 728, 52

\bibitem[\protect\citeauthoryear{{Leitherer} et~al.}{{Leitherer}
  et~al.}{1999}]{leitherer99}
{Leitherer}, C., et~al. 1999, \apjs, 123, 3

\bibitem[\protect\citeauthoryear{{Liang} et~al.}{{Liang}
  et~al.}{2006}]{liang06}
{Liang}, Y.~C., {Yin}, S.~Y., {Hammer}, F., {Deng}, L.~C., {Flores}, H.,  \&
  {Zhang}, B. 2006, \apj, 652, 257

\bibitem[\protect\citeauthoryear{{Limousin} et~al.}{{Limousin}
  et~al.}{2007}]{limousin07}
{Limousin}, M., et~al. 2007, \apj, 668, 643

\bibitem[\protect\citeauthoryear{{Lin} et~al.}{{Lin} et~al.}{2009}]{lin09}
{Lin}, H., et~al. 2009, \apj, 699, 1242

\bibitem[\protect\citeauthoryear{{Madau}}{{Madau}}{1995}]{madau95}
{Madau}, P. 1995, \apj, 441, 18

\bibitem[\protect\citeauthoryear{{Maeder}}{{Maeder}}{1992}]{maeder92}
{Maeder}, A. 1992, \aap, 264, 105

\bibitem[\protect\citeauthoryear{{Mannucci} et~al.}{{Mannucci}
  et~al.}{2010}]{mannucci10}
{Mannucci}, F., {Cresci}, G., {Maiolino}, R., {Marconi}, A.,  \& {Gnerucci}, A.
  2010, \mnras, 408, 2115

\bibitem[\protect\citeauthoryear{{McGaugh}}{{McGaugh}}{1991}]{mcgaugh91}
{McGaugh}, S.~S. 1991, \apj, 380, 140

\bibitem[\protect\citeauthoryear{{Nagao}, {Maiolino}, \& {Marconi}}{{Nagao}
  et~al.}{2006}]{nagao06}
{Nagao}, T., {Maiolino}, R.,  \& {Marconi}, A. 2006, \aap, 459, 85

\bibitem[\protect\citeauthoryear{{Noterdaeme} et~al.}{{Noterdaeme}
  et~al.}{2012}]{noterdaeme12}
{Noterdaeme}, P., et~al. 2012, \aap, 540, A63

\bibitem[\protect\citeauthoryear{{Osterbrock}}{{Osterbrock}}{1989}]{osterbrock89}
{Osterbrock}, D.~E. 1989, {Astrophysics of gaseous nebulae and active galactic
  nuclei} (University Science Books, Mill Valley, CA)

\bibitem[\protect\citeauthoryear{{Pagel} et~al.}{{Pagel}
  et~al.}{1979}]{pagel79}
{Pagel}, B.~E.~J., {Edmunds}, M.~G., {Blackwell}, D.~E., {Chun}, M.~S.,  \&
  {Smith}, G. 1979, \mnras, 189, 95

\bibitem[\protect\citeauthoryear{{Pagel} et~al.}{{Pagel}
  et~al.}{1992}]{pagel92}
{Pagel}, B.~E.~J., {Simonson}, E.~A., {Terlevich}, R.~J.,  \& {Edmunds}, M.~G.
  1992, \mnras, 255, 325

\bibitem[\protect\citeauthoryear{{Pei}, {Fall}, \& {Hauser}}{{Pei}
  et~al.}{1999}]{pei98}
{Pei}, Y.~C., {Fall}, S.~M.,  \& {Hauser}, M.~G. 1999, \apj, 522, 604

\bibitem[\protect\citeauthoryear{{Petitjean}, {Ledoux}, \&
  {Srianand}}{{Petitjean} et~al.}{2008}]{petitjean08}
{Petitjean}, P., {Ledoux}, C.,  \& {Srianand}, R. 2008, \aap, 480, 349

\bibitem[\protect\citeauthoryear{{Pettini} et~al.}{{Pettini}
  et~al.}{2010}]{pettini10}
{Pettini}, M., et~al. 2010, \mnras, 402, 2335

\bibitem[\protect\citeauthoryear{{Pettini} et~al.}{{Pettini}
  et~al.}{1997}]{pettini97}
{Pettini}, M., {King}, D.~L., {Smith}, L.~J.,  \& {Hunstead}, R.~W. 1997, \apj,
  478, 536

\bibitem[\protect\citeauthoryear{{Pettini} \& {Pagel}}{{Pettini} \&
  {Pagel}}{2004}]{pettini04}
{Pettini}, M.,  \& {Pagel}, B.~E.~J. 2004, \mnras, 348, L59

\bibitem[\protect\citeauthoryear{{Pettini} et~al.}{{Pettini}
  et~al.}{2002}]{pettini02}
{Pettini}, M., {Rix}, S.~A., {Steidel}, C.~C., {Adelberger}, K.~L., {Hunt},
  M.~P.,  \& {Shapley}, A.~E. 2002, \apj, 569, 742

\bibitem[\protect\citeauthoryear{{Pettini} et~al.}{{Pettini}
  et~al.}{2000}]{pettini00}
{Pettini}, M., {Steidel}, C.~C., {Adelberger}, K.~L., {Dickinson}, M.,  \&
  {Giavalisco}, M. 2000, \apj, 528, 96

\bibitem[\protect\citeauthoryear{{Pettini} et~al.}{{Pettini}
  et~al.}{2008}]{pettini08b}
{Pettini}, M., {Zych}, B.~J., {Steidel}, C.~C.,  \& {Chaffee}, F.~H. 2008,
  \mnras, 385, 2011

\bibitem[\protect\citeauthoryear{{Pilyugin} \& {Thuan}}{{Pilyugin} \&
  {Thuan}}{2005}]{pilyugin05}
{Pilyugin}, L.~S.,  \& {Thuan}, T.~X. 2005, \apj, 631, 231

\bibitem[\protect\citeauthoryear{{Postman} et~al.}{{Postman}
  et~al.}{2012}]{postman12}
{Postman}, M., et~al. 2012, \apjs, 199, 25

\bibitem[\protect\citeauthoryear{{Quider} et~al.}{{Quider}
  et~al.}{2009}]{quider09}
{Quider}, A.~M., {Pettini}, M., {Shapley}, A.~E.,  \& {Steidel}, C.~C. 2009,
  \mnras, 398, 1263

\bibitem[\protect\citeauthoryear{{Quider} et~al.}{{Quider}
  et~al.}{2010}]{quider10}
{Quider}, A.~M., {Shapley}, A.~E., {Pettini}, M., {Steidel}, C.~C.,  \&
  {Stark}, D.~P. 2010, \mnras, 402, 1467

\bibitem[\protect\citeauthoryear{{Raiter}, {Fosbury}, \&
  {Teimoorinia}}{{Raiter} et~al.}{2010}]{raiter10}
{Raiter}, A., {Fosbury}, R.~A.~E.,  \& {Teimoorinia}, H. 2010, \aap, 510, A109

\bibitem[\protect\citeauthoryear{{Richard} et~al.}{{Richard}
  et~al.}{2011}]{richard11}
{Richard}, J., {Jones}, T., {Ellis}, R., {Stark}, D.~P., {Livermore}, R.,  \&
  {Swinbank}, M. 2011, \mnras, 126

\bibitem[\protect\citeauthoryear{{Rigby} et~al.}{{Rigby}
  et~al.}{2011}]{rigby11}
{Rigby}, J.~R., {Wuyts}, E., {Gladders}, M.~D., {Sharon}, K.,  \& {Becker},
  G.~D. 2011, \apj, 732, 59

\bibitem[\protect\citeauthoryear{{Savaglio}, {Fall}, \& {Fiore}}{{Savaglio}
  et~al.}{2003}]{savaglio03}
{Savaglio}, S., {Fall}, S.~M.,  \& {Fiore}, F. 2003, \apj, 585, 638

\bibitem[\protect\citeauthoryear{{Schaerer} et~al.}{{Schaerer}
  et~al.}{2011}]{schaerer11}
{Schaerer}, D., {Hayes}, M., {Verhamme}, A.,  \& {Teyssier}, R. 2011, \aap,
  531, A12

\bibitem[\protect\citeauthoryear{{Schlegel}, {Finkbeiner}, \&
  {Davis}}{{Schlegel} et~al.}{1998}]{schlegel98}
{Schlegel}, D.~J., {Finkbeiner}, D.~P.,  \& {Davis}, M. 1998, \apj, 500, 525

\bibitem[\protect\citeauthoryear{{Shapley} et~al.}{{Shapley}
  et~al.}{2003}]{shapley03}
{Shapley}, A.~E., {Steidel}, C.~C., {Pettini}, M.,  \& {Adelberger}, K.~L.
  2003, \apj, 588, 65

\bibitem[\protect\citeauthoryear{{Shapley} et~al.}{{Shapley}
  et~al.}{2006}]{shapley06}
{Shapley}, A.~E., {Steidel}, C.~C., {Pettini}, M., {Adelberger}, K.~L.,  \&
  {Erb}, D.~K. 2006, \apj, 651, 688

\bibitem[\protect\citeauthoryear{{Shaw} \& {Dufour}}{{Shaw} \&
  {Dufour}}{1995}]{shaw95}
{Shaw}, R.~A.,  \& {Dufour}, R.~J. 1995, \pasp, 107, 896

\bibitem[\protect\citeauthoryear{{Stark} et~al.}{{Stark}
  et~al.}{2008}]{stark08}
{Stark}, D.~P., {Swinbank}, A.~M., {Ellis}, R.~S., {Dye}, S., {Smail}, I.~R.,
  \& {Richard}, J. 2008, \nat, 455, 775

\bibitem[\protect\citeauthoryear{{Stasi{\'n}ska}}{{Stasi{\'n}ska}}{2005}]{stasinska05}
{Stasi{\'n}ska}, G. 2005, \aap, 434, 507

\bibitem[\protect\citeauthoryear{{Steidel} et~al.}{{Steidel}
  et~al.}{1996}]{steidel96}
{Steidel}, C.~C., {Giavalisco}, M., {Pettini}, M., {Dickinson}, M.,  \&
  {Adelberger}, K.~L. 1996, \apjl, 462, L17

\bibitem[\protect\citeauthoryear{{Swinbank} et~al.}{{Swinbank}
  et~al.}{2007}]{swinbank07}
{Swinbank}, A.~M., {Bower}, R.~G., {Smith}, G.~P., {Wilman}, R.~J., {Smail},
  I., {Ellis}, R.~S., {Morris}, S.~L.,  \& {Kneib}, J.-P. 2007, \mnras, 376,
  479

\bibitem[\protect\citeauthoryear{{Teplitz} et~al.}{{Teplitz}
  et~al.}{2000}]{teplitz00}
{Teplitz}, H.~I., et~al. 2000, \apjl, 533, L65

\bibitem[\protect\citeauthoryear{{Thurston}, {Edmunds}, \& {Henry}}{{Thurston}
  et~al.}{1996}]{thurston96}
{Thurston}, T.~R., {Edmunds}, M.~G.,  \& {Henry}, R.~B.~C. 1996, \mnras, 283,
  990

\bibitem[\protect\citeauthoryear{{van Zee} \& {Haynes}}{{van Zee} \&
  {Haynes}}{2006}]{vanzee06}
{van Zee}, L.,  \& {Haynes}, M.~P. 2006, \apj, 636, 214

\bibitem[\protect\citeauthoryear{{Vanzella} et~al.}{{Vanzella}
  et~al.}{2010a}]{vanzella10b}
{Vanzella}, E., et~al. 2010a, \apj, 725, 1011

\bibitem[\protect\citeauthoryear{{Vanzella} et~al.}{{Vanzella}
  et~al.}{2010b}]{vanzella10a}
{Vanzella}, E., et~al. 2010b, \aap, 513, A20

\bibitem[\protect\citeauthoryear{{Vanzella} et~al.}{{Vanzella}
  et~al.}{2012}]{vanzella12}
{Vanzella}, E., et~al. 2012, \mnras, 424, L54

\bibitem[\protect\citeauthoryear{{Verhamme} et~al.}{{Verhamme}
  et~al.}{2008}]{verhamme08}
{Verhamme}, A., {Schaerer}, D., {Atek}, H.,  \& {Tapken}, C. 2008, \aap, 491,
  89

\bibitem[\protect\citeauthoryear{{Verhamme}, {Schaerer}, \&
  {Maselli}}{{Verhamme} et~al.}{2006}]{verhamme06}
{Verhamme}, A., {Schaerer}, D.,  \& {Maselli}, A. 2006, \aap, 460, 397

\bibitem[\protect\citeauthoryear{{Vernet} et~al.}{{Vernet}
  et~al.}{2011}]{vernet11}
{Vernet}, J., et~al. 2011, \aap, 536, A105

\bibitem[\protect\citeauthoryear{{Villar-Mart{\'{\i}}n}, {Cervi{\~n}o}, \&
  {Gonz{\'a}lez Delgado}}{{Villar-Mart{\'{\i}}n}
  et~al.}{2004}]{villar-martin04}
{Villar-Mart{\'{\i}}n}, M., {Cervi{\~n}o}, M.,  \& {Gonz{\'a}lez Delgado},
  R.~M. 2004, \mnras, 355, 1132

\bibitem[\protect\citeauthoryear{{Villar-Mart{\'{\i}}n}
  et~al.}{{Villar-Mart{\'{\i}}n} et~al.}{1999}]{villar-martin99}
{Villar-Mart{\'{\i}}n}, M., {Fosbury}, R.~A.~E., {Binette}, L., {Tadhunter},
  C.~N.,  \& {Rocca-Volmerange}, B. 1999, \aap, 351, 47

\bibitem[\protect\citeauthoryear{{Weaver} \& {Woosley}}{{Weaver} \&
  {Woosley}}{1993}]{weaver93}
{Weaver}, T.~A.,  \& {Woosley}, S.~E. 1993, \physrep, 227, 65

\bibitem[\protect\citeauthoryear{{Wuyts} et~al.}{{Wuyts}
  et~al.}{2012}]{wuyts12}
{Wuyts}, E., {Rigby}, J.~R., {Sharon}, K.,  \& {Gladders}, M.~D. 2012, \apj,
  755, 73

\bibitem[\protect\citeauthoryear{{Yamada} et~al.}{{Yamada}
  et~al.}{2012}]{yamada12}
{Yamada}, T., {Matsuda}, Y., {Kousai}, K., {Hayashino}, T., {Morimoto}, N.,  \&
  {Umemura}, M. 2012, ArXiv e-prints (1203.3633)

\bibitem[\protect\citeauthoryear{{Yuan} \& {Kewley}}{{Yuan} \&
  {Kewley}}{2009}]{yuan09}
{Yuan}, T.,  \& {Kewley}, L.~J. 2009, \apjl, 699, L161

\end{thebibliography}
\end{document}